\documentclass[journal]{IEEEtran}
\ifCLASSINFOpdf
  \usepackage[pdftex]{graphicx}
  \graphicspath{{../pdf/}{../jpeg/}}
  \DeclareGraphicsExtensions{.pdf,.jpeg,.png}
\else
\fi

\usepackage{amsmath}

\usepackage{subfig}
\usepackage{amssymb}
\newtheorem{theorem}{Theorem}
\usepackage{algorithm}
\usepackage{algpseudocode}

\usepackage{tikz}
\usetikzlibrary{shapes,arrows}
\usepackage{authblk}

\begin{document}
\title{Non-Linear Signal Processing methods for UAV detections from a Multi-function X-band Radar.}

\author{Mohit Kumar, Keith Kelly}
\affil{Agile RF Systems LLC, CO, USA}

\maketitle

\begin{abstract}
	This article develops the applicability of non-linear processing techniques such as Compressed Sensing (CS), Principal Component Analysis (PCA), Iterative Adaptive Approach (IAA) and  Multiple-input-multiple-output (MIMO) for the purpose of enhanced UAV detections using portable radar systems. The combined scheme has many advantages and the potential for better detection and classification accuracy. Some of the benefits are discussed here with a phased array platform in mind, the novel portable phased array Radar (PWR) by Agile RF Systems (ARS), which offers quadrant outputs. CS and IAA both show promising results when applied to micro-Doppler processing of radar returns owing to the sparse nature of the target Doppler frequencies. This shows promise in reducing the dwell time and increase the rate at which a volume can be interrogated. Real-time processing of target information with iterative and non-linear solutions is possible now with the advent of GPU-based graphics processing hardware. Simulations show promising results.
\end{abstract}



\section{Introduction}
\IEEEPARstart{T}{he} main goal of CS is to use optimization methods to recover a sparse signal from a small number of non-adaptive measurements. The radar measurements can be viewed as sparse in both time and Doppler space and are possibly sampled at sub-Nyquist rates, which breaks the relationship between the number of samples acquired and the perfect recovery of radar parameters like delay, velocity, and target angle. The recovery of essential micro-Doppler signatures from the UAV target through the sparse representation of the signal in the frequency domain and following optimization of the sparse signal's $l_{1}$ norm using CS can improve the classification accuracy of various UAV targets. Additionally, MIMO-based virtual aperture formation can impart a better spatial resolution for the small spatial footprint UAV targets. IAA is another Doppler resolution enhancement technique that is considered in this article and it shows a promising application for UAV detections with few pulses. Prior to CS, filtering is accomplished using PCA-based decomposition into eigen sub-spaces to get rid of clutter contamination principally due to sidelobes pointing towards the ground. We develop a unified theory for the applicability of these non-linear processing methods and show their enhancements for better UAV detection using simulations. A common theoretical framework is developed for ease of understanding and applicability of these techniques.\par
\begin{figure}
	\centering
	\includegraphics[width=3in]{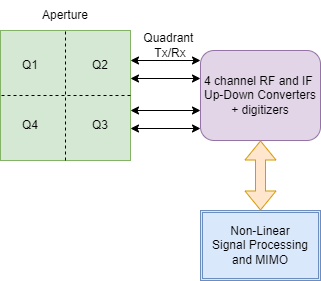}
	\caption{A conceptual representation of a MIMO quadrant phased array for PWR Weather radar system.}
	\label{fig_blkdia}
\end{figure}

For the US Air Force, Agile RF Systems (ARS) has finished developing a portable weather radar (PWR) system built on phased arrays and a four-quadrant architecture. It can be mounted on a roof or tower. It has a sealed radome that provides wind, rain, snow, hail, and sand protection. The CS, IAA and PCA methods elaborated in this article are with reference to this phased array design. The Figure \ref{fig_blkdia} depicts the conceptual representation of the various sub-sections of this phased array radar. The data from the quadrant-based four-phased array centers can be processed by the signal processor and backend processing algorithms implemented in servers. This radar is based on quadrant-level processing to implement a 4-channel MIMO architecture. This article makes use of this hardware platform to demonstrate non-linear processing which also has a quadrant-wise aperture for MIMO related enhancements.

Recent advances in computational methods and increased computing capacity for real-time radar operations have greatly increased the use of non-linear processing in radars and communication. For radar applications, the design complexity is typically higher, and a variety of computational techniques can be used to achieve the desired properties. \cite{mohit2020}. Today's modern phased array radars are able to switch beams faster and are based on inertia-less electronic phase programmability for observing different directions. With such a rapid observation capacity is needed a software framework that can extract information from the least number of acquired samples and pulses, aiding in reducing the dwell time of radar in a specific direction. This ultimately results in overall increase at the rate at which targets can be revisited or surveilled. This article combines the power of non-linear CS, IAA and PCA to make this advancement into the next generation of radar processing and develops a theoretical understanding with modeling of signal, clutter and noise spaces for non-linear processing.\par 
Taking a peek into CS, if $\textbf{x}$ is a sparse vector, we can try recovering it from the knowledge of observation vector $\textbf{y}$ by solving the following optimization problem:
\begin{equation}\label{opt_prob}
arg\; \mathop{min}_{x} ||\textbf{x}||_{0}\;\;\;\; subject\; to\; \textbf{y} = \Theta \textbf{x}
\end{equation}
This search is, however, NP-hard and can be replaced by its closest convex norm, the $l_{1}$ norm \cite{candes2014}. The equation above can thus be reformulated as:
\begin{equation}\label{opt_prob_1}
arg\; \mathop{min}_{x} ||\textbf{x}||_{1}\;\;\;\; subject\; to\; \textbf{y} = \Theta \textbf{x}
\end{equation}
where $\Theta$ is the reconstruction matrix. This condition is influenced by the incoherence of the matrix (the sensing matrix), as well as the sparsity of the initial vector $\textbf{x}$ \cite{candes2014}. The literature offers a number of solutions to this optimization issue. To locate the sparse approximation of the incoming signal $\textbf{x}$ in a dictionary or matrix $\psi$, basis pursuit is used in CS. The Dantzig selector, basis pursuit denoising (BPDN), total variation (TV) minimization-based denoising, etc. are additional commonly used formulations for reliable data recovery from noisy measurements \cite{rani2018}. The squared $l_{2}$-norm of the error between the reconstructed signal $\textbf{y}$ and the sparse signal $\hat{x}$ in the case of BPDN should be less than or equal to $\epsilon$ for the obtained solution.
\begin{equation}\label{opt_prob1}
arg\; \mathop{min}_{x} ||\textbf{x}||_{1}\;\;\;\; subject\; to\; ||\textbf{y} - \Theta \textbf{x}||_{2}^{2} \leq \epsilon
\end{equation}
We can also solve BPDN in its Lagrangian form, which is an unconstrained optimization problem and can be rewritten as:
\begin{equation}\label{opt_prob2}
\hat{x} = arg\; \mathop{min}_{x} \lambda||\textbf{x}||_{1}\; + ||\textbf{y} - \Theta \textbf{x}||_{2}^{2}.
\end{equation}
The primal-dual interior-point technique and fixed-point continuation are two well-known algorithms that have been applied to the aforementioned equation. Algorithms for linear programming, such as the simplex algorithm known as BP-simplex and the interior-point algorithm known as BP-interior, can also be used to solve the optimization problem in equation \ref{opt_prob1}. These are solvers for convex problems. \par
In this article, we seek to develop formulations of the MIMO, CS and PCA operations with PWR as the platform for UAV detections which is a novel combination of non-linear processing techniques not explored before in literature. The data from the four sub-aperture channels are processed by the signal processor and backend processing algorithms implemented in servers kept in an enclosed thermally controlled chamber as part of PWR radar hardware. To enable MIMO and CS-related enhancements, the signal processor implemented in the radar server would need modifications to incorporate MIMO, CS and PCA-related data processing. \par
These non-linear processing methods would eventually lead to the development of low-cost, power-efficient, and small-size radar systems that can scan faster and acquire larger volumes than traditional systems. Here we briefly present the evolution of these methods. Many previous works on CS methods allow recovery of sparse, under-sampled signals from random linear measurements \cite{eldar2012}. In \cite{mishali2011}, authors present Xampling as a sub-Nyquist framework for signal acquisition and processing of signals in a union of subspaces. We are not utilizing Xampling for analog-to-digital conversion. All processing techniques are after the Nyquist rate ADC conversions in fast time. \cite{zhu2020} has used CS to enhance micro-Doppler signatures of drones however, what is lacking is a common framework of understanding and evaluating other non-linear methods like IAA that is presented in our article and how IAA compares against CS in terms of performance. In \cite{gong2022}, an optimal dwell time is evaluated for effectiveness to capture at least one full rotation of the blades. They have commented on the total dwell time required but they don't discuss the sampling rate requirement over the dwell time. The article \cite{eldar2012} serves as a good introduction to and a survey about compressed sensing. In \cite{eldar2014} authors analyze the number of samples required for perfect recovery under noiseless conditions. They have devised a good theoretical framework which we have extended to PCA and IAA under clutter and noise conditions. In \cite{rani2018}, authors have summarized a whole set of optimization routines that can be used to reconstruct a signal using CS. Authors in \cite{candes2014} developed the beginning of a mathematical theory of super-resolution. They illustrated that you can super-resolve point sources with infinite precision i.e. recover the exact locations and amplitudes by solving a simple convex optimization problem, which can essentially be reformulated as a semi-definite program. This holds provided the distance between the sources meet a certain criteria. The article \cite{sira2006} talks about a method that exploits the difference in the statistics of the returns from sea clutter and the target to improve detection performance. On the other hand, we selected PCA as the dominant approach to remove clutter echoes by suppressing clutter eigen-vectors and also removing few noise eigen-vectors to enhance SNR. In \cite{wang2016}, authors discuss Subspace space–time adaptive processing (STAP) algorithms
to eliminate clutter. However, estimation of clutter sub-space is a severe limitation. \par

This article explores using CS and IAA-based reconstruction of micro-Doppler for small UAV targets from fewer pulses, such that we do not lose micro-Doppler characteristics for the detection and classification of these targets. Traditionally using Fourier transform on these fewer pulses will degrade the resolution to such an extent that the nearby micro-Doppler features cannot be identified. This aspect is simulated using CS and IAA performance versus FFT-based reconstruction and the benefits can be readily observed. The MIMO formulation is also presented which aids in better spatial resolution indeed needed to support the accurate localization of these small targets.

\begin{figure*}
	\centering
	\includegraphics[width=5in]{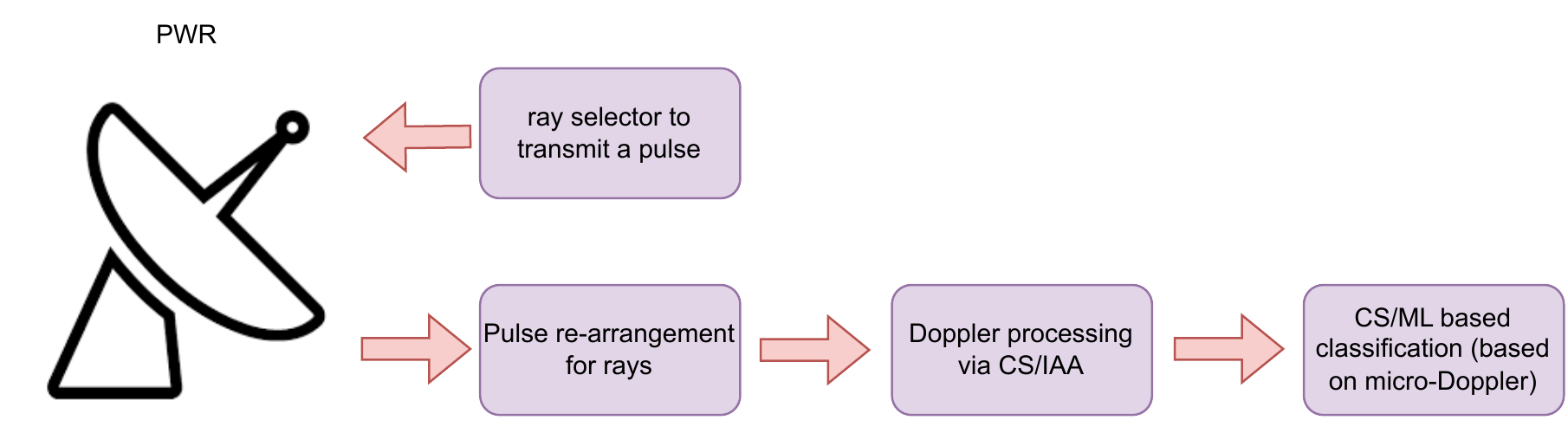}
	\caption{Simple illustration of the processing system.}
	\label{fig_bd}
\end{figure*}
Figure \ref{fig_bd} gives the basic conceptual processing steps needed for building up this system. As would be evident later that for CS-based recovery from a minimum number of pulses, the pulses must be randomly transmitted in different elevation states (in the case of PWR radar) thus we need a random ray (direction) selector to send out a pulse. At the receiver, we would need to segregate all the pulses together for a ray and process along the slow time (pulse) axis for Doppler super-resolution. IAA however, doesn't have this requirement which can be one of its advantages as compared to CS. Uniform sampling, in case of IAA, also aids in PCA-based clutter suppression which would not work for the non-uniformly sampled received echo. In that case, we need to wait for all the pulses to go out in all directions (rays) for PCA and CS to start. In case of IAA, however, we can start as soon as one ray (direction) echoes have been received.\par
There is an urgent need for faster scanning for a drone detection radar system because these small objects are highly agile and maneuvering. A really fast update is required to surveillance and track the full space for drones and swarms of drones to keep an eye on their ever-changing activities and strategies. We need counter UAS systems like PWR equipped with very fast scan strategies using very few pulses in a direction and still able to recover high-resolution Doppler features from detected drones. These non-linear processing techniques would aid PWR in achieving this goal.

\subsection{Portable Weather Radar (PWR)}
\begin{figure}
	\centering
	\subfloat[\centering The Radar on its tower at the CSU-CHILL radar test facility in Greeley.]{\includegraphics[width=5cm]{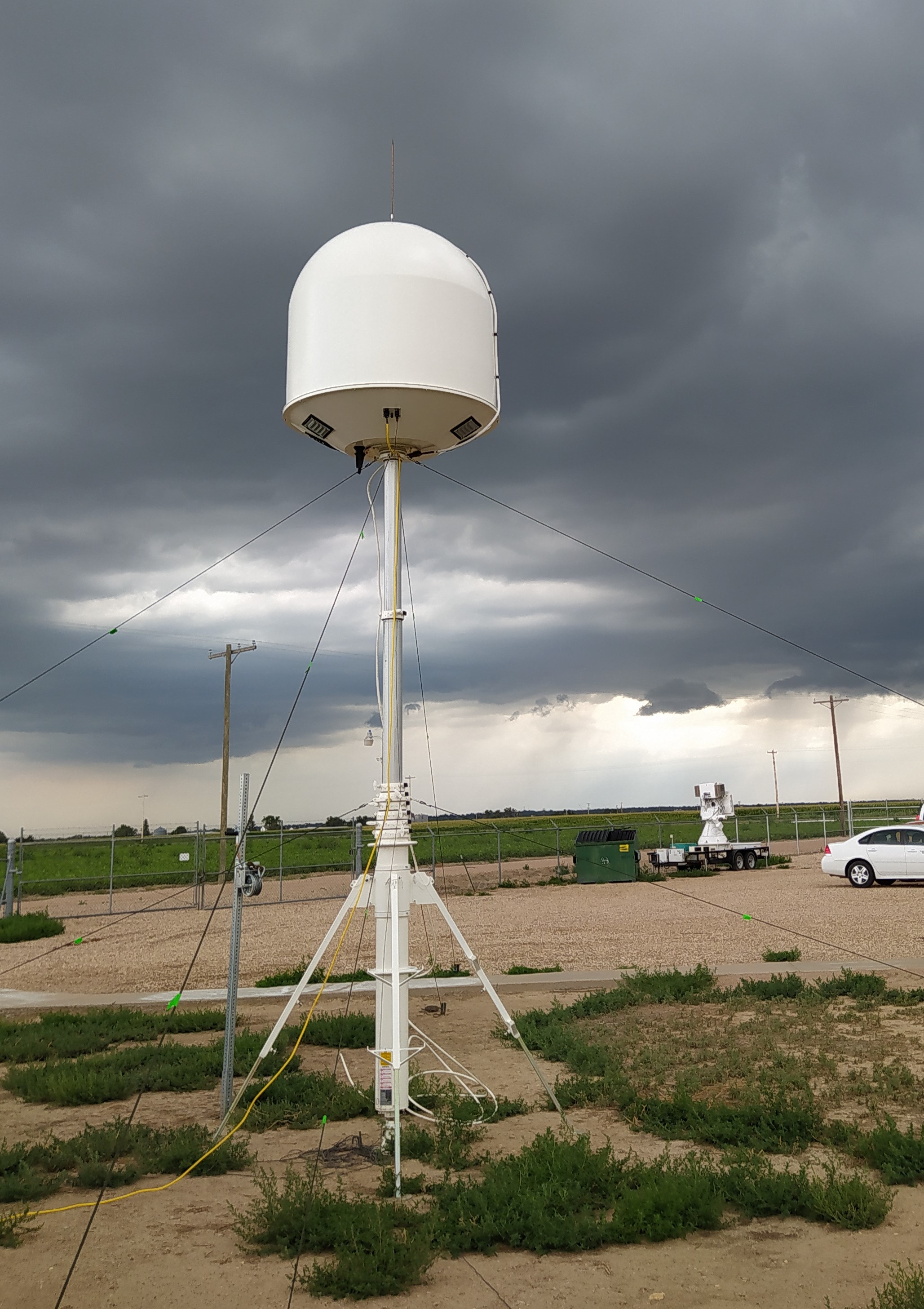} }
	\qquad
	\subfloat[\centering The RTS setup with horns on a tower pointing towards radar in a far field.]{\includegraphics[width=7cm]{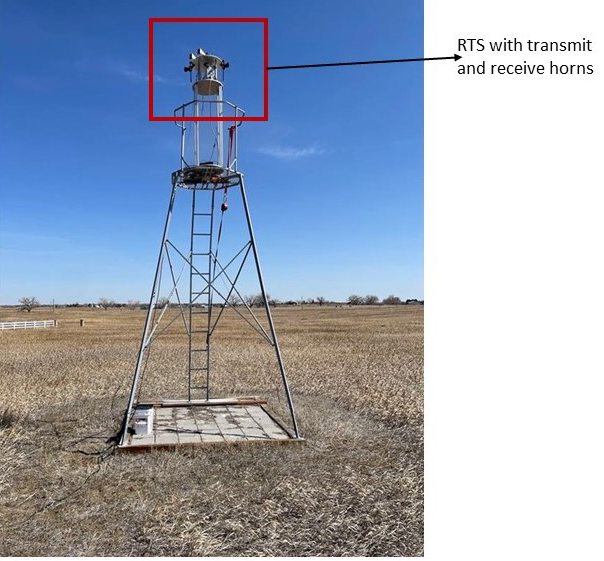} }
	\qquad
	\subfloat[\centering The beamshape obtained.]{\includegraphics[width=7cm]{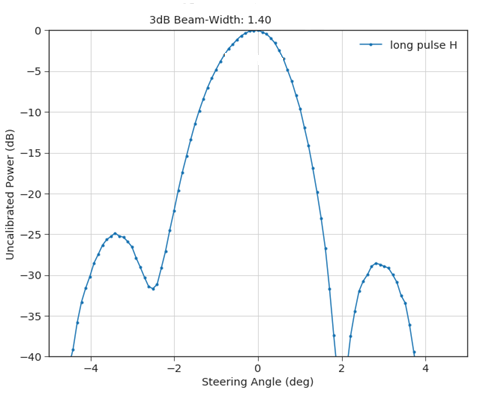} }
	\caption{Radar and RTS Setup \cite{kelly2022}}%
	\label{fig_radar}%
\end{figure}

All the techniques discussed in this article are being developed with reference to ARS PWR weather radar sensor. PWR is a flexible and agile radar due to the phase spin architecture and central Radar System Controller (RSC) \cite{kelly2022}. This radar is based on a phased array design and is inherently very different from parabolic dish antenna radars like D3R \cite{mohit2017,mohit2018}. The radar located at the Greeley radar test facility is shown in Figure \ref{fig_radar}(a). The phase gradient that the phased array controller uses is coordinated by the RSC, along with the motor control for azimuth positioning and rotation. The FPGA logic in Software defined radio (SDR) has a programmable register interface that enables the RSC to change a broad range of operational radar parameters. The RSC uses alternate horizontal and vertical polarization to allow transmit pulses with a Linear Frequency Modulated Chirp (LMFC) waveform in the SDR. The Host Processor in the local cabinet receives the filtered radar returns from the multichannel receive hardware and processes them there.

The beamforming network and phased array antenna for PWR underwent extensive testing. Array calibration and aperture beam pattern data were collected to confirm expected aperture performance. An internal Radar Target Simulator (RTS) was created to test the complete functionality of the radar system with the calibrated aperture. The RTS was positioned for this test in far-field of the aperture (Figure \ref{fig_radar} (b)). The PWR waveform was received by RTS, a digital time delay was applied, and the result was transmitted back to the PWR while it was still in receive mode. An accurate estimate of the combined beam pattern and all four quadrants was confirmed by determining the peak value from the returned waveform in PWR at each elevation. The two-way combined H-pol antenna pattern is shown in Figure \ref{fig_radar} (c) measured with the help of RTS. The two-way pattern sidelobes are approximately 25 dB below the mainlobe peak power and the 3dB beamwidth measured is 1.4 deg confirming good phase/time alignment of all quadrant channels in the combined pattern. With a similar setup of RTS, using MIMO Coherent implementation, we expect to measure ~0.9 to 1 deg of 3 dB beamwidth, improvements coming through the four-quadrant based MIMO signal processing.\par

\begin{figure*}
	\centering
	\includegraphics[width=5in]{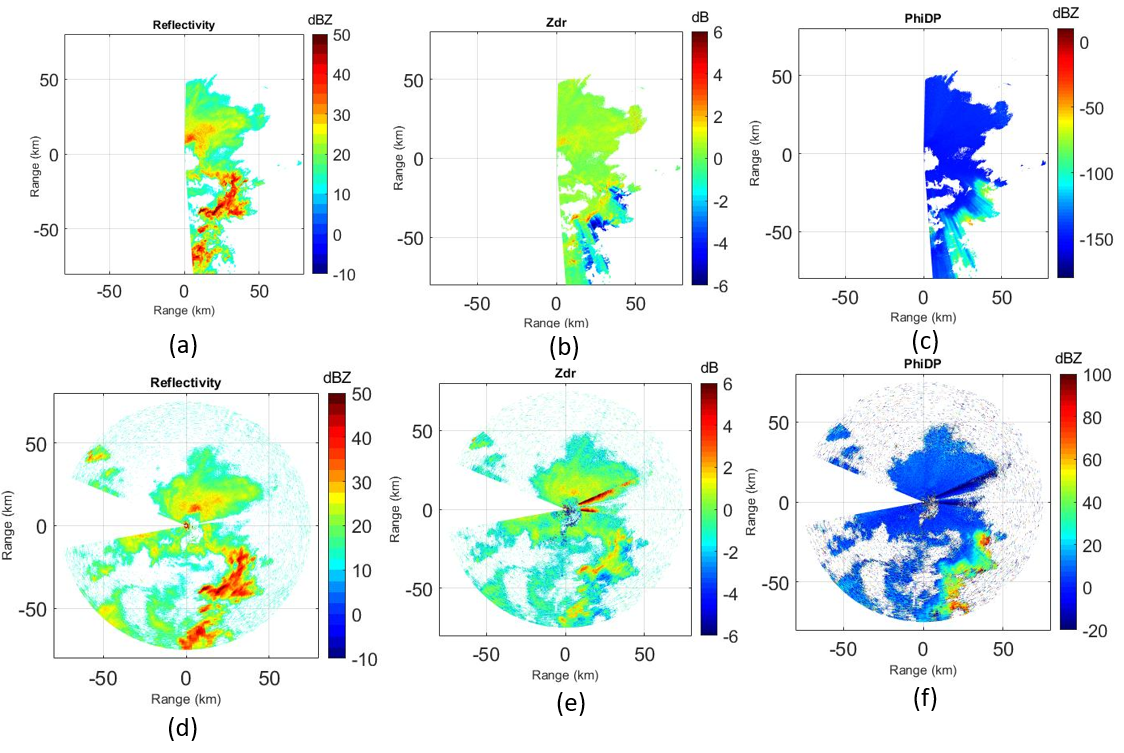}
	\caption{The different polar products being generated by CHILL and PWR radars \cite{kelly2022}.}
	\label{fig_dataComp}
\end{figure*}

To remove any spatial ambiguity, PWR was co-located with the CSU-CHILL radar and concurrently gathered weather observations. This was done to determine the PWR data products' level of quality for SISO-based radar operations. It has been described here so that it can serve as a standard by which to compare the efficacy of MIMO. Data comparisons between these two radars were carried out while PWR rotated at a constant quarter RPM and CHILL transmitted in the eastern region for the same 14 elevation states. Let's examine one of the light rain cases that both radars recorded on May 31, 2022. Figure \ref{fig_dataComp}(a) and (d) show the reflectivity field for CHILL and PWR radars respectively while (b) and (e) shows the comparison of differential reflectivity between the two. Figure \ref{fig_dataComp}(c) and (f) show the differential phase being encountered going through the storm from these radars. The top plots are from the CHILL radar and the bottom ones are from PWR. The CHILL radar was scanning only the eastern sector while the PWR did the whole 360-degree coverage. Both radars observed 14 elevation states and the 2 deg elevation state is shown in the figures. Several of the bright thunderstorm features that both radars picked up in the southeast can be seen distinctly in these figures. All of the level 2 products were subjected to this comparison. With nine times larger antenna dimensions than PWR, we can easily see the CHILL radar's high spatial resolution. With MIMO Coherent processing, the spatial resolution of PWR is anticipated to improve without a physical size increase in an effort to resolve the weather storm features better.\par
PWR can be very easily configured for sensing both weather and drone targets. This is the hardware platform developed at ARS currently being used for weather sensing. Using a separate processing chain shown in Figure \ref{fig_bd}, we can easily expand its capabilities for drone target detection using non-linear processing. It is fully capable of MIMO aperture extension because of its four quadrants transmit and receive channels and on similar terms, because of its software-defined capabilities in terms of beam agility and waveforms, it is an ideal platform for testing out non-linear techniques.

Spectrogram and smoothed-pseudo Wigner-Ville distribution are two time-frequency representation techniques that have been widely used to analyze drone micro-Doppler signatures. Furthermore, a number of classification methods based on micro-Doppler signatures have been reported for classifying drones of various sizes, types, and loads, as well as drones and people, dogs, and birds. The radar antennas in real-world ground-based surveillance radar systems must scan rapidly to cover a large spatial area of up to 360°. This implies that the radar beam's dwell period on any given target is quite short (which is, usually, a few tens of milliseconds). Thus, when adopting the conventional fast Fourier transform (FFT) for Doppler processing, the radar Doppler resolution is very poor and the accurate micro-Doppler signatures of drones are difficult to discriminate.\par

\section{Methods}
\subsection{PCA and CS Formulation for micro-Doppler enhancement}
Radar echoes from drones can be identified, categorized, and tracked using Micro-Doppler. A spinning blade is a feature of the majority of drones, including single-rotor, quadrotor, six-rotor, and even hybrid vertical takeoff and landing (VTOL) drones. They are typically active in low-altitude airspace, are small, and fly slowly \cite{musa2019,wellig2018}. The rotating movement of rotating blades can modulate the incident radar wave and produce an additional micro-Doppler on the base of the body Doppler contributed by the flying motion of the drone body. Micro-Doppler signals are thought to be quite useful signatures for radar-based drone detection and classification \cite{gong2022}.\par
The importance of micro-Doppler for drone detections cannot be overstated. Using lengthy FFT sizes in traditional signal processing, drone detections can be more accurately resolved at higher Doppler resolutions. In general, greater Doppler resolution is associated with longer radar dwell times (sending out more pulses for longer FFTs). However, the maximum radar dwell period for a functional radar sensor applies. A practical radar system should be able to track targets more quickly and look quickly in all directions to search the entire volume. The secret to observing such a micro-Doppler is the radar dwell time. The dwell period should be sampled quickly enough to improve the Doppler resolution of our spectral analysis. CS and IAA-based non-linear processing can break this relation of linear dependence of resolution to the number of pulses required to observe micro-Doppler features of drones \cite{eldar2014}.\par
Prior to performing CS/IAA, we begin with PCA to get rid of clutter contamination of the drone echo. The cleaned-up signal can then go through spectral analysis.
\subsection{PCA decomposition of Clutter}
Consider a radar transceiver, similar to \cite{eldar2014} that transmits a pulse train:
\begin{equation}\label{eq_signalMod}
\textbf{x}_{T}(t) = \sum_{p=0}^{P-1} h(t-p\tau),\;\; 0 \leq t \leq P\tau.
\end{equation}
consisting of P equally spaced pulses $h(t)$. The pulse-to-pulse delay $\tau$ is referred to as the PRI, and it's reciprocal $1/\tau$ is the PRF. The entire span of the signal in equation \ref{eq_signalMod} is called the coherent processing interval (CPI). Let L Doppler-producing drone targets make a scene. The pulses travel back to the transceiver after reflecting off the L targets. Three parameters are used to describe each target: a complex amplitude $\alpha_{l}$ that is proportional to the target's radar cross-section (RCS), a Doppler radial frequency $\upsilon_{l}$ that is proportional to the target-radar closing velocity, and a time delay $\tau_{l}$ that is proportional to the target's distance from the radar. We can write the received signal as:

\begin{equation}\label{eq_recsignalMod}
x(t) = \sum_{p=0}^{P-1}\sum_{l=0}^{L-1}\alpha_{l} h(t - \tau_{l} - p\tau)e^{-j\upsilon_{l}p\tau}.
\end{equation}
It might be convenient to express the signal as a sum of single frames:
\begin{equation}\label{eq_recsignalMod1}
x(t) = \sum_{p=0}^{P-1}x_{p}(t)
\end{equation}
where
\begin{equation}\label{eq_recsignalMod2}
x_{p}(t) = \sum_{l=0}^{L-1}\alpha_{l} h(t - \tau_{l} - p\tau)e^{-j\upsilon_{l}p\tau}.
\end{equation}
This is the case when the target can be characterized using a single velocity, $\upsilon_{l}$, however, in case of micro-Doppler frequencies we have a band of frequencies around the main body Doppler component comprising the micro-Doppler ($\Delta\upsilon_{l}$) as:
\begin{equation}\label{eq_recsignalMod2_1}
x_{p}(t) = \sum_{l=0}^{L-1}\sum_{i=0}^{I-1}\alpha_{l} h(t - \tau_{l} - p\tau)e^{-j(\upsilon_{l}+\Delta\upsilon_{i})p\tau}.
\end{equation}
$I$ are the number of micro-Doppler components.\par
Additionally, in practice, this signal is contaminated with noise and clutter:
\begin{equation}\label{eq_recsignalMod3}
x(t) = \sum_{p=0}^{P-1}[x_{p}(t) + \omega_{p}(t) + C_{p}(t)].
\end{equation}
where $\omega(t)$ is a zero mean wide-sense stationary random signal with auto-correlation $r_{\omega}(s) = \sigma^{2}\delta(s)$ and $C(t)$ is the clutter component. A synonymous equation quantized in time would be:
\begin{equation}\label{eq_recsignalMod3_1}
x(n) = \sum_{p=0}^{P-1}[x_{p}(n) + \omega_{p}(n) + C_{p}(n)].
\end{equation}
In order to decrease its effect on micro-Doppler features, removing the clutter component is necessary. The equation \ref{eq_recsignalMod3_1} can be thought to be composed of signal, noise and clutter sub-spaces. Let the mean values of $\textbf{x}_{p}(n)$ be $\mu_{p}$. Then the mean subtracted received signal can be written as:

\begin{equation}\label{eq_recsignalMod4}
x(n) = \sum_{p=0}^{P-1}(x_{p}(n) - \mu_{p}).
\end{equation}
Forming the auto-correlation matrix $\textbf{R}_{xx}$ of $\textbf{x}(n)$ and performing SVD decomposition on it yields,
\begin{equation}\label{eq_recsignalMod4}
\textbf{R}_{xx} = \textbf{U}\textbf{S}\textbf{V}^{T}.
\end{equation}
Sorting out the eigenbasis vectors in $\textbf{U}$ in descending order, we get the largest principal components in the received signal. If clutter is supposed to be the dominant return signal component, we can set the corresponding eigenvector corresponding to the largest eigenvalue in $\textbf{S}$ to zeros. The rest of the received signal $x_{p}(t)$ is projected to remaining eigenvectors and sum them up as follows to reconstruct signal and noise:
\begin{equation}\label{eq_recsignalMod4}
x_{recons}(n) = \sum_{p=0}^{P-1}\sum_{d=0}^{D-1}u_{d}(n)x_{p}(n).
\end{equation}
$u_{d}'s$ span the eigenvector space comprising of signal and noise sub-spaces minus the clutter sub-space as that eigenvector is not part of this space. The signal and noise sub-spaces are orthogonal to each other. We can also reduce noise power by considering only a few noise eigenvectors and adding them up to the signal sub-space. This might improve SNR. A demonstration of this is part of the simulations section. A point worth noting is that we can figure out clutter power by averaging out the received signal whose mean will give an estimate of clutter power centered at DC. Based on this, we can come to know which eigenvector should be removed to nullify the clutter sub-space, if clutter is not the most dominant echo in the received signal.\par
If we compare PCA with MTI clutter filter, we can observe that MTI removes the clutter component while low-frequency micro-Doppler components can also get completely suppressed, which would decrease the distinction of micro-Doppler features and finally influence the classification accuracy of drone targets. After the clutter signal has been suppressed, we go to CS step for enhancing micro-Doppler features. Micro-Doppler spectral lines can have better distinction when they are CS processed.\par
Apart from the primary signal Doppler, a few micro-Doppler lines, and clutter, which have high values, the majority of the entries in the spectral domain of drone targets are zeros or low values. Only the primary Doppler and a few high spectral lines caused by micro-Doppler may remain after removing clutter. In order to improve drone classification and identification using fewer pulse samples, CS may be able to provide high-resolution Doppler components for such a sparse signal. If we use fewer pulses to give the same resolution as with, say, 10 times the number of pulses, then we are effectively reducing the dwell time on the target and can potentially spin faster as in the case of PWR. This faster scanning radar can track and do multiple functions at the same time which may mean portable systems like PWR, is able to accomplish weather surveillance, track, and surveillance UAVs.
\subsection{CS-based enhancement of Doppler Space}
The clutter-suppressed received signal can now be processed by CS to better resolve the micro-Doppler frequencies with  relatively fewer random measurements. The premise is CS would be able to provide a higher resolution Doppler space with few samples as against conventional FFT processing which would need a sufficiently larger number of measurements or pulses to give the same resolution as CS. The first stage of CS is multiplying the random measurement matrix, $\psi(n)$ with $x(n)$:
\begin{equation}\label{eq_recsignalMod3_2}
y(n) = \sum_{p=0}^{P-1}\psi_{p}(n)[x_{p}(n) + \omega_{p}(n) + C_{p}(n)].
\end{equation}
where $\psi_{p}(n) \in \mathbb{R}^{MxN} \;or\; \mathbb{C}^{MxN}$ and $y(n) \in \mathbb{R}^{M} \;or\; \mathbb{C}^{M}$. The number of measurements taken is much lesser than the length of the input signal, i.e., $M<<N$. To further reduce the number of measurements which are necessary for perfect reconstruction, the measurement matrix must be incoherent with the basis in which the signal is sparse. The inputs to the reconstruction algorithm are the measurement vector $y(n)$ and reconstruction matrix $\Theta$ where,
\begin{equation}
\Theta = \psi\xi \in \mathbb{R}^{MxN} \; or\; \mathbb{C}^{MxN}.
\end{equation}
$\xi$ is the basis vector of the space where $x(n)$ is sparse. Thus $x(n)$ can be written as:
\begin{equation}
x(n) = \sum_{p=0}^{P-1}s_{p}(n)\xi_{p}(n).
\end{equation}
$s\in \mathbb{R}^{N}$ is the sparse coefficient vector of length N. The optimization problem expressed as $l_{1}$ norm (for reconstruction) can thus be expressed as:
\begin{equation}
\hat{s} = arg\; \mathop{min}_{s} ||\textbf{s}||_{1}\;\;\;\; subject\; to\;  \Theta \textbf{s} = \textbf{y}
\end{equation}
The estimate of x(n), i.e., $\hat{x}$ can be obtained from $\hat{s}$ by taking its inverse transform. Some of the other types of this same optimization problem with noise included and a lagrangian form of the above equation were discussed in the paragraphs preceding equations \ref{opt_prob1} and \ref{opt_prob2}. \par
It is shown in the literature \cite{eldar2014} that for a noise-free case, the estimation of parameters $(\alpha_{l}, \tau_{l}, \upsilon_{l})_{l=0}^{L-1}$ without micro-Doppler frequencies can be recovered using 3L samples using a Xampling framework and assuming Finite Rate of Innovation (FRI) samples. However, with micro-Doppler and the presence of noise, there are likely more samples required for perfect recovery. Simulations do confirm the fact that the number of slow time, pulse measurements required for a higher resolution Doppler reconstruction is sufficiently less so that either the radar can be made to scan faster or it can be made to accomplish multi-functions like weather detections and forecasting too. The software-defined phased array architecture of PWR is ideally suited for drone detection and weather surveillance.
\subsubsection{SNR}
The SNR is linked to the attenuation that the signal receives going through the link which is very obvious based on the CS model. However, noise power can be reduced in the prior step of PCA with fewer noise eigenvectors considered for the reconstruction of the received signal. Certainly, this can improve SNR and it is demonstrated in simulations too.
\subsection{MIMO and CS Framework}
The multi-function PWR radar is capable of MIMO because of its four-quadrant array structure. So we should be able to use quadrant-wise MIMO formulation along with CS. This gives the benefit of virtual array formation without the addition of physical array elements, and also it is cost-effective since each element is not required to have an RF and IF hardware chain associated, and only four channels are sufficient to make use of quadrant MIMO benefits instead of hundreds if not thousands of channels for a full MIMO implementation.\par
\begin{figure}
	\centering
	\includegraphics[width=3in]{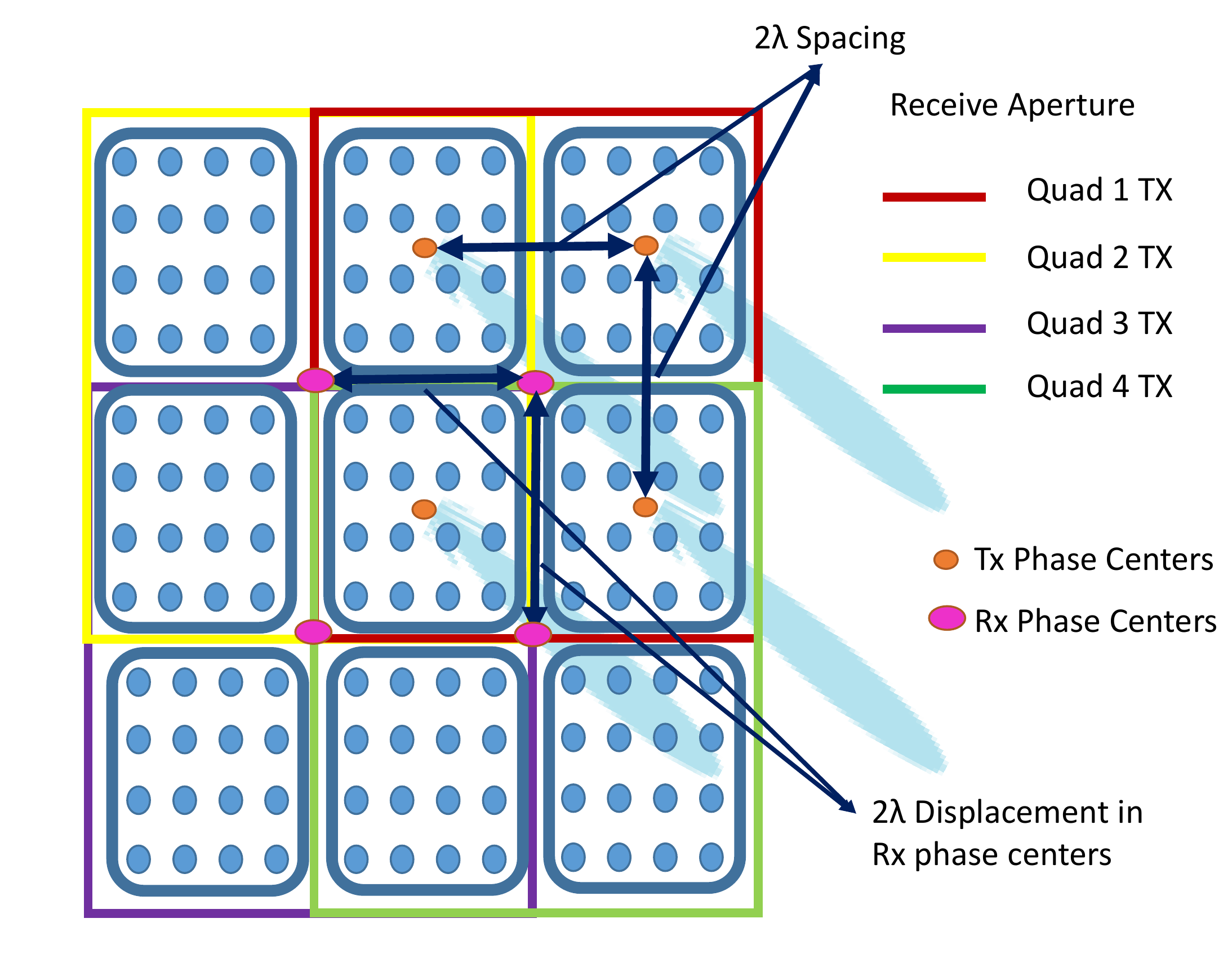}
	\caption{A 8x8 element phased array with multiple transmit phase centers based on the quadrant. The whole array is divided into 4 quadrants \cite{mohit2023}. }
	\label{fig_phasedWeather1}
\end{figure}

The quadrant MIMO system is equivalent to the spatial convolution of the transmit and receive quadrant phase centers and the formation of virtual array elements beyond the physical aperture size. The virtual array dimensions are 1.5 times the physical array (in both axes) as evident from Fig. \ref{fig_phasedWeather1}. Equivalently, this would give the beamwidth reduction by the same factor and the spatial resolution will improve. The PWR system provides a very cost-effective MIMO radar system using a quadrant phased array structure. One of the main challenges of an element-wise MIMO radar is coping with complicated systems in terms of cost, high computational load, and complex implementation, which have been traded very well using quadrant MIMO in PWR radar hardware.
To demonstrate quadrant MIMO processing, we assume transmissions using the same LFM waveforms from all the quadrants, however, we would transmit sequentially from quadrants in a time-multiplexed manner. The data cube received by the quadrants would have to be processed to form the virtual array data cube. Let the data collected when quadrant 1 transmits (from all four receive quadrants) given by:
\begin{equation}
\textbf{Ph}_{r} =
\begin{bmatrix}
\textbf{M}_{11} & \textbf{M}_{12} \\
\textbf{M}_{13} & \textbf{M}_{14}
\end{bmatrix}
\label{eq1}
\end{equation}

Where the first subscript tells the transmit quadrant and the second signifies the receive. The coherent data matrix after all transmissions is given by:
\begin{equation}
\textbf{Vi}_{r} =
\begin{bmatrix}
\textbf{M}_{21} & \textbf{M}_{11}+\textbf{M}_{22}  & \textbf{M}_{12}\\
\textbf{M}_{23}+\textbf{M}_{31} & \textbf{M}_{13}+\textbf{M}_{24}+\textbf{M}_{31}+\textbf{M}_{41}  & \textbf{M}_{14}+\textbf{M}_{42}\\
\textbf{M}_{333} & \textbf{M}_{43}+\textbf{M}_{43}  & \textbf{M}_{44}\\
\end{bmatrix}
\label{eq12}
\end{equation}
This matrix includes 5 additional virtual phase centers corresponding to five additional quadrants to make it a total of 9 quadrants. For PWR, each entry in the column is $12\lambda$ wide and high, the third row and column would give it an extra $12\lambda$ height and width to the receive aperture due to virtual quadrants \cite{mohit2023}.\par
Extending our discussion further about MIMO and CS, let's revisit equation \ref{eq_recsignalMod2} for a sparse scene with L drone targets. The received signal at the $q_{th}$ quadrant after demodulation to baseband for a single frame is in turn given by:
\begin{equation}\label{eq_recsignalMod2_1}
\textbf{x}_{q}(t) = \sum_{l=0}^{L-1}\sum_{m=0}^{M-1}\alpha_{l} h(t - \tau_{l} - p\tau)e^{-j\upsilon_{l}p\tau}e^{-j \beta_{m,q}\varsigma}.
\end{equation}
where $\varsigma=sine(\theta_{l})$ is the azimuth angle of the $l^{th}$ drone target relative to the quadrant $\theta_{l}$. Also note that, $\beta_{m,q} = (\zeta_{q}\xi_{m})(f_{c}\lambda/c + 1)$, $f_{c}$ is the carrier frequency radiated from the quadrant and $\zeta_{q},\xi_{m} \in \textbf{Vi}_{r}$. Again $y(n)$ can be written as:
\begin{equation}\label{eq_recsignalMod3_2}
y(n) = \sum_{p=0}^{P-1}\psi_{p}(n)[x_{p,q}(n) + \omega_{p}(n) + C_{p}(n)].
\end{equation}
and then forming $\textbf{s}$, an sparse coefficient matrix using basis $\iota$ as:
\begin{equation}
x(n) = \sum_{p=0}^{P-1}\sum_{q=0}^{M-1}s_{p,q}(n)\iota_{p,q}(n).
\end{equation}
The optimization problem expressed as $l_{1}$ norm (for reconstruction) can now be expressed as:
\begin{equation}
\hat{s} = arg\; \mathop{min}_{s} ||\textbf{s}||_{1}\;\;\;\; subject\; to\;  \Theta \textbf{s} = \textbf{y}
\end{equation}
\begin{theorem}
	The minimal number of transmit times the number of receive channels required for perfect recovery of L targets in noiseless settings is $ \geq 2L$ with a minimal number of $\geq 2L$ samples per receiver and $\geq 2L$ pulses per transmitter \cite{eldar2012}.
\end{theorem}
This is true for Xampling and an FRI framework used in conjunction with CS. In PWR, quadrant MIMO offers four transmit and nine receive channels (four physical and five virtual quadrants), thus $L=18$ drone targets can be resolved in a CPI or dwell time $\tau$. For this recovery, 36 samples are needed per pulse and 36 pulses are needed per transmitter quadrant for perfect recovery of Doppler for these L targets. This arithmetic is different for a noisy link but then most probably we don't have that many drone targets too. Only in the case of drone swarms, we may need to be detecting more targets, however, it would be quite a coincidence to get so many of them in a CPI or dwell (one direction) otherwise. This result also implies that many more targets can be perfectly resolved in DoA sense by using MIMO virtual elements and this framework allows CS theory to be applied for calculating the number of pulses required for a perfect Doppler recovery for all these targets as well.

\subsection{An Iterative Approach to solve the dwell time limitation for a fast scanning drone radar}
The Doppler resolution of the temporal signal can be increased by using the super-resolution algorithms that are frequently used in array processing, such as minimal variance distortionless response (MVDR) and multiple signal classification (MUSIC). To estimate the covariance matrix or carry out eigen analysis, these algorithms typically need a number of signal snapshots. Some algorithms, like MUSIC, require knowing the number of sources up front as well. However, in surveillance radar, the Doppler processing is carried out over the slow-time samples (over pulses) at each range increment. As a result, there is only one available temporal snapshot. It is also unclear how many target Doppler and micro-Doppler sources there will be. Consequently, it is impossible to use the traditional super-resolution methods. Unlike the conventional MVDR and MUSIC algorithms in which many snapshots are required to estimate the covariance matrix, IAA can work
well with only a few or even one snapshot to achieve super-resolution.\par
\begin{algorithm}
	\caption{An iterative algorithm \cite{xue2011} }\label{alg:cap}
	\begin{algorithmic}
		\State $\hat{P}_{k} = \frac{1}{\textbf{a}^{H}\textbf{a}\sum_{p=0}^{P- 1}|\textbf{a}^{H}(f_{k})y(n)|^{2}}$
		\While {!converge}
		\State $\textbf{R} =\textbf{A}(f)\textbf{P}\textbf{A}^{H}(f)$
		\For {k = 1,2,...,K}
		\State $\hat{s}_{k} = \frac{\textbf{a}^{H}(f_{k})\textbf{R}^{-1}\textbf{y}}{\textbf{a}^{H} (f_{k})\textbf{R}^{-1}\textbf{a}(f_{k})} \; \; n=1,2,...,N$.
		\State $\hat{P}_{k} = 1/N \sum_{n=0}^{N-1} |\hat{s}_{k}(n)|^2$.
		\EndFor
		\EndWhile
	\end{algorithmic}
\end{algorithm}
The formulation of this method is elaborated next. It is similar to the one highlighted in \cite{xue2011}. The basis steering vectors defined on the grid points that either have the frequency present or not span the output space of the Doppler processor. We can, henceforth, write the outcome of the Doppler process as:
\begin{equation}
\textbf{y} = \textbf{A}(f)\textbf{s} + \omega + \textbf{C}. 
\end{equation}
where $\omega(n)$ is a zero mean wide-sense stationary random signal with auto-correlation $r_{\omega}(s) = \sigma^{2}\delta(s)$ and C(n) is the clutter component. $A(f) =[\textbf{a}(f_{1}) \textbf{a}(f_{2}) ... \textbf{a}(f_{k})]$ is $PxK$ dimension where P is the number of pulses and K is the number of finite points in the Doppler grid. \textbf{s} is a vector of the amplitudes of frequencies at the grid locations $k=1,2,...,K$. The clutter and noise matrix can be defined as:
\begin{equation}
\textbf{Q}(f_{k}) = \textbf{R}-P_{k}\textbf{a}(f_{k})\textbf{a}^{H}(f_{k}). 
\end{equation}
$\textbf{R} =\textbf{A}(f)\textbf{P}\textbf{A}^{H}(f)$ is the auto-correlation matrix of the input and P is a K×K diagonal matrix, whose diagonals
$P_{k} = |s_{k}|^{2}, k = 1,2,...,K$ contains the powers at each Doppler frequency on the Doppler grid. The cost function is given by:
\begin{equation}
\Xi = (\textbf{y} - s_{k}\textbf{a}(f_{k}))^{H}\textbf{Q}^{-1}(\textbf{y} - s_{k}\textbf{a}(f_{k})). 
\end{equation}
Minimizing the cost function with respect to $s_{k}$ gives \cite{xue2011}:
\begin{equation}
\hat{s}_{k} = \frac{\textbf{a}^{H}(f_{k})\textbf{R}^{-1}\textbf{y}}{\textbf{a}^{H}(f_{k})\textbf{R}^{-1}\textbf{a}(f_{k})}. 
\end{equation}
Since the iteration requires $\textbf{R}$, which depends on the unknown powers, it must be implemented as an iterative approach. The initialization can be done by letting $\textbf{R}$ equal to the identity matrix $\textbf{I}_{P}$. The steps are shown in Algorithm \ref{alg:cap}.
Both IAA and CS are capable to enhance the Doppler resolution with fewer pulses, however, CS needs a random pulse transmission within the dwell time. It relies on non-uniform sampling within the dwell time. The pulse time left vacant because a pulse cannot be transmitted in CS can be used for transmitting pulses in other directions though, however, it can make radar operations complex. On the other hand, IAA works with a uniform sampling of the dwell time.
\subsection{Simulations and Discussions}
\begin{figure*}
	\centering
	\includegraphics[width=5in]{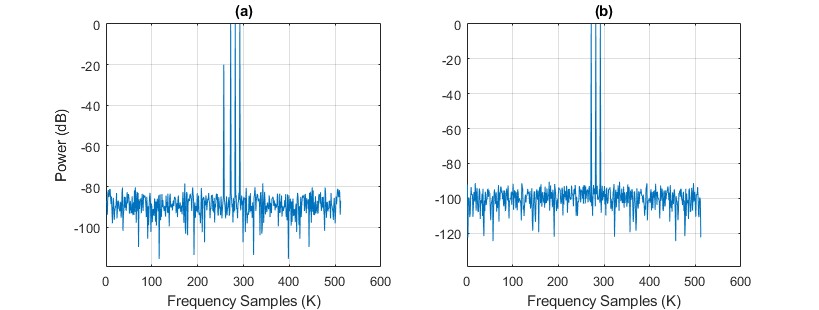}
	\caption{Clutter removal using PCA. (a) is with clutter centered at DC and (b) clutter removed with no harm to nearby micro-Doppler components. }
	\label{fig_pca}
\end{figure*} 
\begin{figure}
	\centering
	\includegraphics[width=4in]{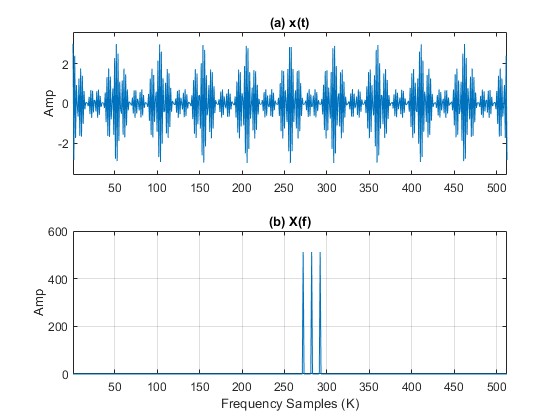}
	\caption{The original signal characteristics for a CS-based micro-Doppler reconstruction. }
	\label{fig_cs1}
\end{figure}
\begin{figure}
	\centering
	\includegraphics[width=3in]{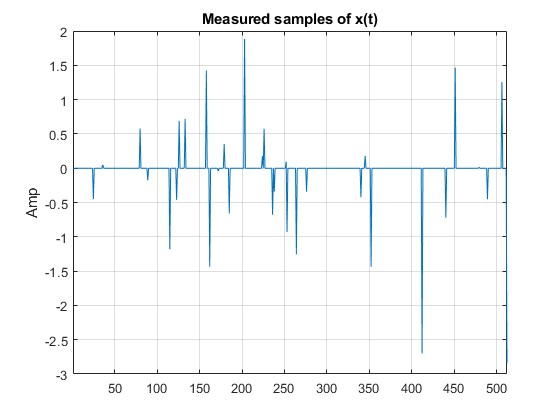}
	\caption{The samples that are picked randomly and CS based reconstruction is applied. }
	\label{fig_cs2}
\end{figure}

 
In this section, we show the feasibility and practicality of non-linear methods discussed in earlier sections by simulations using PWR radar parameters and features. We begin by simulating a few micro-Doppler frequencies making an echo of length 512 samples. This comprises a main Doppler echo from the base motion of the UAV and there are micro-Doppler from the rotary motion of the blade movement modulating the primary echo signal. The signal is corrupted by clutter from the elevation sidelobes of PWR simulated as zero Doppler component being added up to the received echo. The PCA formulation described in an earlier section is used next for the removal of clutter sub-space and reconstruction of the time domain signal for further processing. Current methods including MTI, CLEAN, etc can not realize the real-time removal of ground clutter without suppressing nearby micro-Doppler components. That is why PCA is adopted to remove the clutter components in the echo signal. Figure \ref{fig_pca} (a) depicts clutter centered at DC and the micro-Doppler components and the main Doppler signal. If clutter is always a dominant signal in the received echo, then it is pretty easy to remove the highest valued eigenvector from the SVD decomposition, however, if it is not, then we need to figure out the DC power by averaging out the samples and looking at average power. This should be able to give us an estimate of the eigenvector which has similar power levels as the mean power. After that is made sure, we can remove it from our reconstruction process to get rid of clutter. Next, this signal is processed using CS. The time domain and the frequency domain of the echo for CS processing are shown in Fig. \ref{fig_cs1} (a) and (b). Random 32 samples (out of 512) are picked from $x(t)$ (Fig. \ref{fig_cs2}) and the recovery of the sparse frequency domain is accomplished from these random samples using $l_{1}$ minimization. 

\begin{figure}
	\centering
	\includegraphics[width=4in]{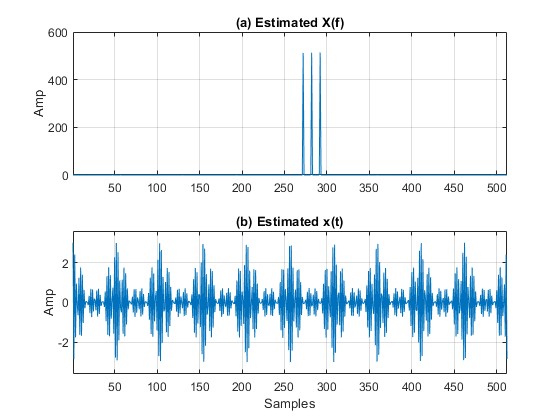}
	\caption{The reconstructed frequency domain and time domain. }
	\label{fig_cs4}
\end{figure}

The recovered high-resolution frequency and time domain samples from the lower dimensional signal are shown in Figure \ref{fig_cs4} (a) and (b). 

\begin{figure}
	\centering
	\includegraphics[width=3in]{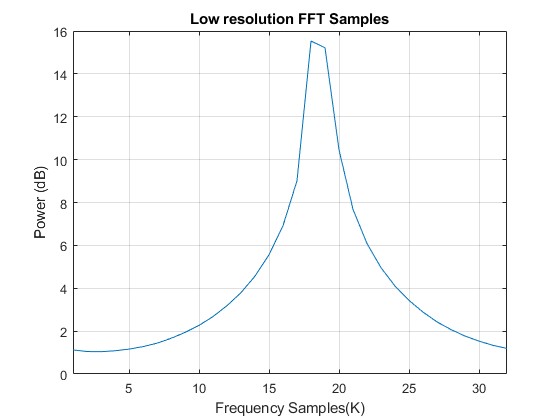}
	\caption{The Fourier transform of the lower-dimensional signal. }
	\label{fig_fft1}
\end{figure}

Exact recovery of the higher dimensional signal is possible due to sparsity in the frequency domain. The frequency analysis of the lower dimensional signal is shown in Fig. \ref{fig_fft1} which is the Fourier transform of the first 32 samples from the sequence of the original 512 samples. The loss of resolution is evident and the modulations due to micro-Doppler cannot be observed. This would lead to faulty classification results for the UAV type and detection of UAV based on certain micro-Doppler features. 

It is to be noted that CS needs random $K=32$ samples from a set of $N=512$ echo samples. The $N$ can be considered here to be the number of pulses where we reduced it to $K << N$. Thus only $K$ pulses are sufficient to reconstruct the micro-Doppler features of the UAV echo which can easily reduce the dwell time and overall scan time of the radar. These pulses would need to be randomly transmitted in the larger dwell time of $N$ pulses, scanning other elevation states to cover up the total volume in the case of PWR for example. With this, the $N$ pulses would be transmitted at random in different elevation states and they can be all clubbed together in the signal processor. This is a little complex and would need the beam switching time every pulse instead of every dwell in case of uniform sampling of pulses. The IAA however, doesn't rely on non-uniform random sampling and is simulated next.

\begin{figure}[t]
	\centering
	\includegraphics[width=3.8in]{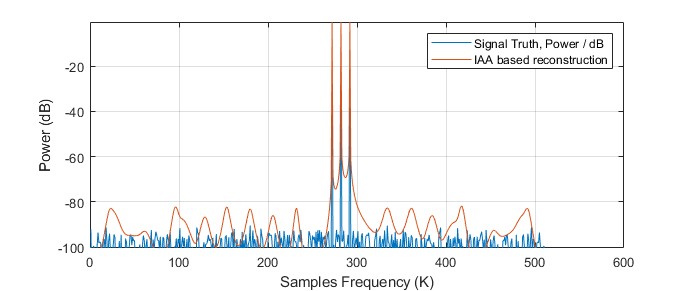}
	\caption{Estimation of frequency response using IAA.}
	\label{fig_sim4}
\end{figure}  

Figure \ref{fig_sim4} shows the same setup of micro-Doppler frequencies. Instead of the CS-based reconstruction, we are using IAA for recovery. It is shown that super-resolution can be achieved using a single snapshot of data samples. The IAA iteration was setup simulating three peaks of the micro-Doppler base echo and the modulations from the rotation of the blades of the quadcopter similar to the one for CS and then using $K(32 pulses) << N(512 pulses)$ that are much lesser than the higher dimensions (512) in the frequency domain required to reconstruct a higher resolution frequency response. We can clearly see that the sidelobes are very low for IAA-based reconstruction. To achieve a similar level of sidelobe performance, a very aggressive taper would be needed for FFT-based recovery that would lead to quite a bit of SNR loss. IAA can achieve higher Doppler resolution with even higher SNR than the conventional FFT with more pulses for one snapshot. With this, micro- Doppler signatures of drones can be clearly discriminated, which greatly benefits the subsequent classification of drones. The higher Doppler resolution can also help to separate the slow-moving targets and the ground clutter at zero Doppler, and therefore improve the detectability of
multi-rotor drone which usually moves very slowly. Using IAA can avoid the taper loss in FFT-based Doppler processing and the overall radar detection performance for all targets is also improved. If we compare it with CS technique, CS works with random non-uniform sampling that is unconventional and as we said earlier about it's applicability to PWR, we converged on a scheme in which different elevation states would be selected at random for transmission of a pulse, that scheme is complicated as compared to uniformly sampled IAA. Having said this, it is worth noting that CS can be extended to fast time sampling using Xampling and FRI principles so that lower sampling ADCs are sufficient for below the Nyquist rate sampling of fast time signals. Hence both schemes have their own pros and cons and should be judiciously used.\par

\section{Conclusion}
We looked into and simulated PCA for clutter mitigation explored CS and IAA for micro-Doppler and spectral retrievals, MIMO for spatial estimation of drone UAV targets. A unified theoretical framework was developed that stitches together all these non-linear areas towards drone micro-Doppler enhancement for detections from phased array PWR multi-function radar sensor. Both IAA and CS were found to be very useful to recover micro-Doppler drone features so that those targets can be efficiently detected and classified using fewer pulses than conventional FFT processing. The drawbacks and applicability of each one of these techniques were discussed.



\end{document}